# Geometric and Signal Strength Dilution of Precision (DoP) Wi-Fi

**Soumaya ZIRARI\*, Philippe CANALDA and François SPIES**

[1] **Computer Science Laboratory of the University of Franche-Comté, France**
Numerica, 1 cours, Louis Leprince Ringuet, 25200 Montbéliard
*E-Mail: soumaya.zirari@univ-fcomte.fr*

[2] **Computer Science Laboratory of the University of Franche-Comté, France**
Numerica, 1 cours, Louis Leprince Ringuet, 25200 Montbéliard
*E-Mail: philippe.canalda@univ-fcomte.fr*

[3] **Computer Science Laboratory of the University of Franche-Comté, France**
Numerica, 1 cours, Louis Leprince Ringuet, 25200 Montbéliard
*E-Mail: francois.spies@univ-fcomte.fr*

**Abstract**
The democratization of wireless networks combined to the emergence of mobile devices increasingly autonomous and efficient lead to new services. Positioning services become overcrowded. Accuracy is the main quality criteria in positioning. But to better appreciate this one a coefficient is needed. In this paper we present Geometric and Signal Strength Dilution of Precision (DOP) for positioning systems based on Wi-Fi and Signal Strength measurements.

*Keywords:* Wireless LAN, Radio position measurement, Indoor radio communication.

## 1. Introduction

The world population is currently growing which implies a remarkable increase in buildings and skyscrapers. These are obstacles for *Global Navigation Satellite Systems* (GNSS) such as the *Global Positioning System* (GPS). New networks have emerged (UMTS, GSM, ...) which does not help to reduce the impact of interferences. These factors among others contribute to the GPS [1] up to 20 meter loss in accuracy especially in urban and peri-urban environments.

During the last ten years the number of users of the IEEE 802.11x community has known a remarkable growth and a new positioning solution based on Wi-Fi was born. Some positioning algorithms guaranty an accuracy of 5 meters such as RADAR[2], Viterbi-like algorithm [3], *Friis and Reference Based Hybrid Model* [4] (FRBHM) [5].

The GPS is limited in given environments and Wi-Fi is becoming a viable positioning method. The authors think that the Wi-Fi network can be adapted by learning from the GPS.

In this paper, we present a mathematical approach of a new version of the known GPS *Dilution of Precision* [6] which is more adapted to the Wi-FI networks and use other elements to estimate the precision. We also present a model that allows to estimate the precision based on criteria other than the geometric one only.
The third section presents and analyzes some results.

## 2. GEOMETRIC CRITERIA

The evolution of the IEEE 802.11 standard fulfil more and more the constraints allowing the improvement of its efficiency in large and more complex environments.

The efficiency of such networks is measured by different criteria. Some of those criteria are focused on the network geometry, others on the throughput [7] or on the interference [8].

*A. Gondran and al.* [9] provide a geometric indicator for WLAN planning. This indicator is based on the study of the covered area by a Basic Service Set (BSS), where a cell relative to one antenna is a set of pixels associated to a given base station. The cell $C$ is defined by:

$$c = \{b_{i,j} / F_{i,j} \quad q\} \qquad (1)$$

IJCSI

Where is $b_{i,j}$ the pixel of coordinates $i, j$ and $F_{i,j}$ is the signal strength received at $b_{i,j}$ exceeding a given quality threshold $q$.

Considering the 2-D space, each pixels have 8 neighbours with the exception of the pixels on space borders. *Mabed and al.* [10] define the geometrical criteria as bellow:

$$G'(C) = \frac{\sum_{b_{i,j} \in C} V(b_{i,j})}{8 \times |C| - 6 \times \sqrt{\pi \times |C|}} \quad (2)$$

*A. Gondran and al.* adapted this formula to 3-D space which can be indoor environment such as buildings.

$$G'(C_k) = \frac{\sum_{b_{i,j,k} \in C_k} V(b_{i,j,k})}{8 \times |C_k| - 6 \times \sqrt{\pi \times |C_k|}} \quad (3)$$

Where $k$ presents the floor.

The geometric indicator regrouping all floor-indicators is defined by the following equation:

$$G_{WLAN}(C) = \sum_{k=1}^{k=K} \frac{|C_k|}{|C|} G'(C_k) \quad (4)$$

Where

$$C = \bigcup_{k=1}^{k=K} C_k$$

## 3. Propagation Models

When the signal transmitted by a transmitter travels in space, it loses its power. Part of the energy of the signal strength is dissipated. The environment where the carrier signal travels and the distance covered have an important impact on the signal attenuation.

Several equations have been developed.

### 3.1 FRIIS

The *Friis* [11] equation is:

$$\frac{P_R}{P_T} = G_R G_T \left(\frac{\lambda}{4\pi d}\right)^2 \quad (5)$$

where :

· $P_R$ and $P_T$ are respectively the Signal Strength (*SS*) received and the *SS* emitted;

· $G_R$ and $G_T$ are respectively the receiver and transmitter antenna gains;

· is the carrier wavelength;

· $d$ is the distance between the receiver and the transmitter.

### 3.2 Interlink Networks

The Interlink Networks [14] approach offers to replace the power 2 in the Friis formula by the power to the 3.5 due to the prompt wave's attenuation in a building because of the high number of obstacles in this one.

The Interlink Networks formula is:

$$\frac{P_R}{P_T} = G_R G_T \left(\frac{\lambda}{4\pi d}\right)^{3.5} \quad (6)$$

where :

- $P_R$ and $P_T$ are respectively the Signal Strength (*SS*) received and the *SS* emitted;
- $G_R$ and $G_T$ are respectively the receiver and transmitter antenna gains;
- is the carrier wavelength;
- $d$ is the distance between the receiver and the transmitter.

### 3.3 SNAP-WPS

*Y. Wang* proves in the paper [15] the possibility to approximate the target position by measuring the signal strength. In fact, the signal attenuation between the transmitter and the receiver allows to determine the mobile position. However, the Friis equation enables to estimate the distance between the receiver and the transmitter in an environment without any obstacles. Thus, Y. Wang suggest an empirical model based on regression. By comparing the residual among different degrees polynomials, he decide that a cubic regressive equation would be adequate for the empirical model $EM^2$:

$$d_i = 0.000198 \; S_i^3 - 0.025 \; S_i^2 + 1.14 \; S_i - 14.8 \quad (7)$$

Where *S* is the signal strength (SS) in dBM, normally is between 15-90 dBM.

### 3.4 Analysis

The results in Table.1 [16] present the comparison between Wi-Fi positioning systems.

Table 1: Comparison Between The Positioning Algorithms

| Positioning System | Mean Error | Standard Deviation |
|---|---|---|
| Friis | 9.86 | 6.3 |
| SNAP-WPS | 8.76 | 5.87 |
| Interlink Networks | 9.58 | 5.11 |
| FBCM | 7.77 | 3.03 |
| Radar | 4.62 | 2.98 |
| FRBHM | 5.98 | 3.22 |

## 2. Contribution

The contribution of this paper consists in presenting a precision of dilution model for wireless networks. This model aims at giving an idea about the position estimation accuracy. This model can be described in three steps:

1- The first step consists in the constitution of a set of all visible access points (Fig. 1). The number of visible access points is one of the decisive elements on the accuracy of a positioning system. Our needs in the number of visible access points depend on the dimension of the positioning system. At least three APs for a two dimension positioning system and at least four APs for a three dimension one. If the number of AP is not sufficient, we set automatically the value of the precision of dilution coefficient as infinite. The optimal value is equal to one.

2- The second step concerns the signal strength of the visible access points (Fig. 2). We assume that access points with a signal strength under a given threshold may induce errors in the position estimation of the target. An access point with a bad signal strength can be near or far from the user. In fact, the signal strength may be attenuated either because of the distance or because of the number of obstacles. If only three access points have a good signal strength (we are in a 3D positioning system) we predict that the coefficient value will be higher.

3- The final step deals with the positioning system architecture geometry *i.e.* the third step verifies if the visible access points are geometrically well distributed with respect to the user. For this step we propose a Wi-Fi DOP *Dilution Of Precision* which is calculated as below.

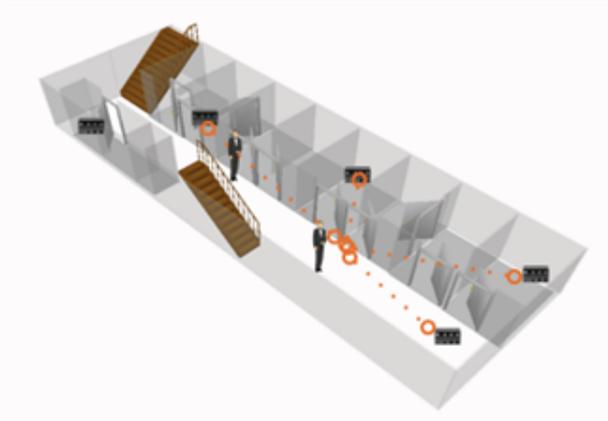

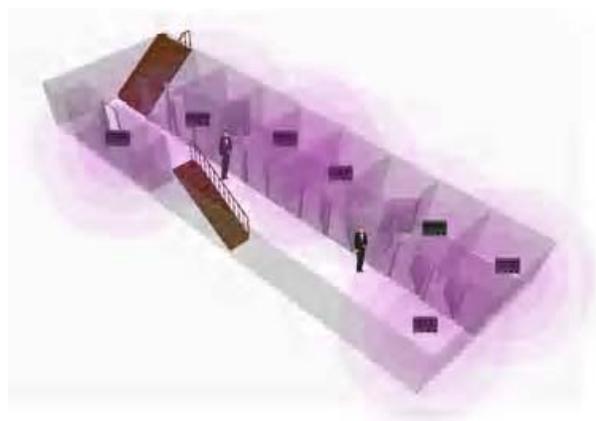

Let us suppose $S_{AP} = N_{AP}$ the number of visible access points. We assume that:

$$S_{AP} = \{AP_1, AP_2, ..., AP_{N_{AP}}\}$$

Where $AP_i$ are the visible access points.
The radius of circle $d_i$ ( $i \in \{1,...,N_{AP}\}$ the number of calculation) is defined by:

$$d_i = \sqrt{(X^{c,i} - X_u)^2 + (Y^{c,i} - Y_u)^2 + (Z^{c,i} - Z_u)^2}$$ (8)

$X^{c,i}, Y^{c,i}, Z^{c,i}$ are the $AP_i$ coordinates and $X_u, Y_u, Z_u$ the user unknown coordinates.

$$\begin{cases} d_1 = \sqrt{(X^{c,1} - X_u)^2 + (Y^{c,1} - Y_u)^2 + (Z^{c,1} - Z_u)^2} \\ d_2 = \sqrt{(X^{c,2} - X_u)^2 + (Y^{c,2} - Y_u)^2 + (Z^{c,2} - Z_u)^2} \\ d_3 = \sqrt{(X^{c,3} - X_u)^2 + (Y^{c,3} - Y_u)^2 + (Z^{c,3} - Z_u)^2} \\ \vdots \\ d_i = \sqrt{(X^{c,4} - X_u)^2 + (Y^{c,4} - Y_u)^2 + (Z^{c,4} - Z_u)^2} \end{cases}$$

We obtain:

(9)

$$\frac{P_R}{P_T} = G_R G_T \left(\frac{\lambda}{4\pi d}\right)^2$$

4.1 Friis equation

The Friis equation [13] as seen before is:

The Friis equation allows us to compute the distance as below:

$$d_i = \frac{\lambda}{4\pi} \sqrt{\frac{P_{T,i} G_R G_{T,i}}{P_{R,i}}}$$ (10)

Where:
$P_{R,i}, P_{T,i}, G_R$ and $G_{T,i}$ are respectively the receiver and $AP_i$ data.

$$\hat{d_i} = \sqrt{(X^{c,i} - \hat{X}_u)^2 + (Y^{c,i} - \hat{Y})^2 + (Z^{c,i} - \hat{Z}_u)^2}$$

The distance $d_i$ can be approximated by a *Taylor expansion*:
(11)

The *Taylor expansion* at the first order is:

$$\Delta d_i = d_i - \hat{d_i} = b_{i,x}\Delta x_u + b_{i,y}\Delta y_u + b_{i,z}\Delta z_u$$

(12)

Where

$$b_{i,x} = \frac{X_{c,i} - X_u}{r_i}, b_{i,y} = \frac{Y_{c,i} - Y_u}{r_i},$$

$$b_{i,z} = \frac{Z_{c,i} - Z_u}{r_i} \text{ and }$$

$$r_i = \sqrt{(X_{c,i} - X_u)^2 + (Y_{c,i} - Y_u)^2 + (Z_{c,i} - Z_u)^2}$$

We obtain:

$$\begin{cases} \Delta d_1 = b_{1,x}\Delta X_u + b_{1,y}\Delta Y_u + b_{1,z}\Delta Z_u \\ \Delta d_2 = b_{2,x}\Delta X_u + b_{2,y}\Delta Y_u + b_{2,z}\Delta Z_u \\ \Delta d_3 = b_{3,x}\Delta X_u + b_{3,y}\Delta Y_u + b_{3,z}\Delta Z_u \\ \vdots \\ \Delta d_i = b_{i,x}\Delta X_u + b_{i,y}\Delta Y_u + b_{i,z}\Delta Z_u \end{cases}$$ (13)

The linear system is:

$$\Delta d = H \Delta X \quad (14)$$

$$\Delta d = \begin{bmatrix} \Delta d_1 \\ \Delta d_2 \\ \Delta d_3 \\ \vdots \\ \Delta d_i \end{bmatrix}$$

We suppose that $P_{T,i}$, $G_R$ and $G_{T,i}$ are fixed parameters. Only $P_{R,i}$ the Signal Strength (*SS*) received from the $AP_i$ is unknown and then estimated.
Thus from the equation (5), we obtain:

$$d_i = \sqrt{\frac{P_{T,i} G_{T,i} G_R}{4}} \sqrt{\frac{1}{P_{R,i}} - \frac{1}{\hat{P}_{R,i}}} \quad (15)$$

We have:

$$H = \begin{bmatrix} b_{1,x} & b_{1,y} & b_{1,z} & 0 \\ b_{2,x} & b_{2,y} & b_{2,z} & 0 \\ b_{3,x} & b_{3,y} & b_{3,z} & 0 \\ & & \vdots & 0 \\ b_{i,x} & b_{i,y} & b_{i,z} & 0 \end{bmatrix}$$

and :

$$DeltaX = \begin{bmatrix} \Delta x_u \\ \Delta y_u \\ \Delta z_u \end{bmatrix}$$

We obtain:

$$C \Delta P_R = H \Delta X \quad (16)$$

Where C is a known matrix equal to:

$$C = \begin{bmatrix} c_1 \\ c_2 \\ c_3 \\ \vdots \\ c_i \end{bmatrix}$$

and $c_i = \sqrt{\frac{P_{T,i} G_{T,i} G_R}{4}}$

$$\Delta P_R = \begin{bmatrix} \Delta P_{R,1} \\ \Delta P_{R,2} \\ \Delta P_{R,3} \\ \vdots \\ \Delta P_{R,i} \end{bmatrix}$$

Where :

$$\Delta P_{R,i} = \sqrt{\frac{1}{P_{R,i}}} - \sqrt{\frac{1}{\hat{P}_{R,i}}}$$

The G matrix is defined by:

$$G = (H^T H)^{-1} \quad (17)$$

The Wi-Fi GDOP follows the equation bellow:

$$DOP = \sqrt{Tr[G]} \quad (18)$$

We conclude from the model that we can estimate the positioning accuracy, and measure the error of the wireless positioning system by analysing the following elements:

- $P_R$ Signal Strength (*SS*) received from the AP which emits a *SS* $P_T$
- $G_R$ the user antenna gain and $G_T$ the AP antenna gain;
- the carrier wavelength;
- The number of visible AP.

4.2 Interlink Networks

The Interlink Networks formula is:

$$\frac{P_R}{P_T} = G_R G_T \left(\frac{\lambda}{4\pi d}\right)^{3.5}$$

The distance is:

$$\Delta d_i = \frac{\lambda \sqrt[3.5]{P_{T,i} G_{T,i} G_R}}{4\pi} \left(\frac{1}{\sqrt[3.5]{P_{R,i}}} - \frac{1}{\sqrt[3.5]{\hat{P_{R,i}}}}\right)$$

(19)

The linear system become equivalent to:

$$C \cdot \Delta P_R = H \cdot \Delta X$$

Where C is a known matrix equal to:

$$C = \begin{bmatrix} c_1 \\ c_2 \\ c_3 \\ \vdots \\ c_i \end{bmatrix}$$

$$c_i = \frac{\lambda \sqrt[3.5]{P_{T,i} G_{T,i} G_R}}{4\pi}$$

and

$$\Delta P_R = \begin{bmatrix} \Delta P_{R,1} \\ \Delta P_{R,2} \\ \Delta P_{R,3} \\ \vdots \\ \Delta P_{R,i} \end{bmatrix}$$

Where :

$$\Delta P_{R,i} = \frac{1}{\sqrt[3.5]{P_{R,i}}} - \frac{1}{\sqrt[3.5]{\hat{P_{R,i}}}}$$

The G matrix is defined by:

$$G = \left[H^T H\right]^{-1}$$

The Wi-Fi GDOP follows the equation bellow:

$$DOP = \overline{Tr[G]}$$

4.3 SNAP-WPS

The distance in SNAP-WPS system is equal to:

$$d_i = 0.000198 \; S_i^3 - 0.025 \; S_i^2 + 1.14 \; S_i - 14.8$$

The linear system is:

$$\Delta S = H \cdot \Delta X \quad (20)$$

## 5. Experiments

Experiments have been carried out to validate our model of precision dilution for wireless networks. *Open Wireless Positioning System* (OWLPS) [17], which is an indoor positioning system, based on the Wi-Fi wireless network, was the positioning system used to calculate the mobile position. The experiments were carried in our laboratory, *Laboratoire d' Informatique de Franche Comté* (LIFC).

5.1 OWLPS Architecture

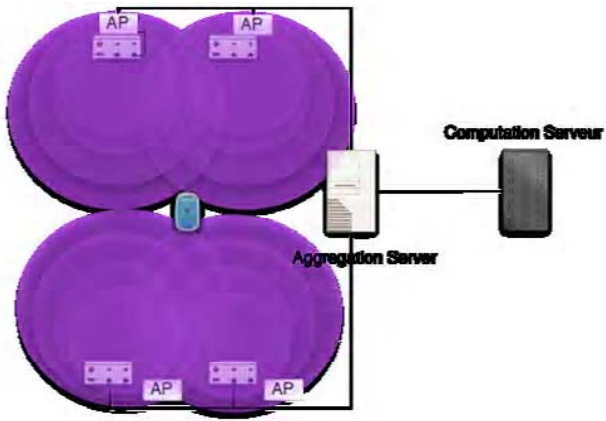

*Open Wireless Positioning System* (OWLPS) implements several positioning techniques and algorithms such as FBCM [4] or FRBHM [5]. The system is Infrastructure-centred, i.e., the mobile asks its position to the infrastructure (see Fig. 3). The main task of the system is to provide an adequate environment to the creation and test of new techniques, propagation models and for the development of hybrid techniques combining existing algorithms.

5.2 The experimentation scenario

As we can see in Fig.4, the experimentation scenario was about a mobile displacements during an interval of time. During all this interval, the user is located through the OWLPS system and the algorithm used for the positioning are Friis, Interlink Networks and FRBHM.

Along all the mobile trajectory, we know the exact mobile coordinates and the estimated one, which allow us to analyse the results. The positioning system is a 3D one.

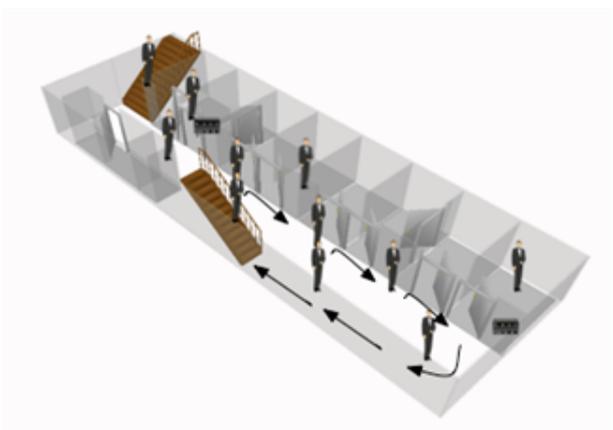

Fig. 3. : The environment of experimentation

5.3 Analysis

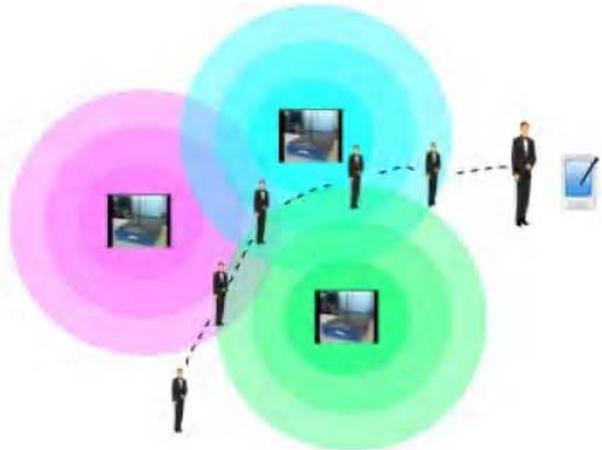

The first experiments were done in order to verify the impact of the number of access points on our model and to check if the Wi-Fi DOP is consistent with this information.

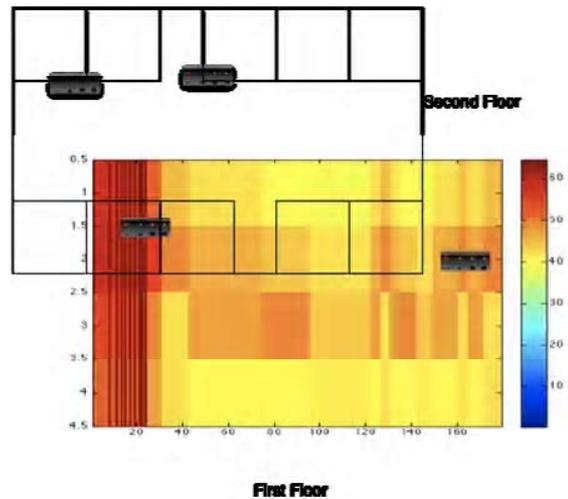

Fig. 6. : The DOP cartography when the mobile is moving in the first floor

The Fig. 6 proves how the Geometric and Signal Strength Dilution of Precision (DoP) Wi-Fi progress with the mobile movement in the first floor of the building.

IJCSI

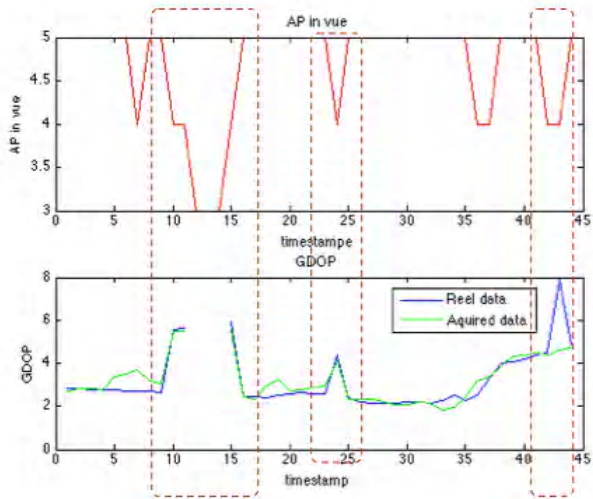

Analysing the results presented in Fig.7, we deduce the DOP fits quite well in terms of number of visible access points. In fact, as shown in Fig.7 when the DOP[ 10,15], the number of visible access points is equal to three, thus the Wi-Fi DOP values reach infinite values.

However, the Wi-Fi DOP values reach good values when the number of visible access points is up to four but we observe some peaks when the number acceptable of access point for 3D positioning system is minimal (*i.e.* four access points).

The second step of our experiments was done in order to verify the impact of the signal strength of each visible access point on our model and in which way this information makes the DOP vary.

Fig. 8 proves that the Wi-Fi DOP is really influenced by the signal strength of the access points. When DOP [10,15] and DOP [35,44], the Wi-Fi DOP values vary from seven to the infinite. If we look at the signal strength for those behaviours we note that the signal can not be received or the signal is too weak. This means that the model can in fact predict the system accuracy by analysing the access point signal strength.

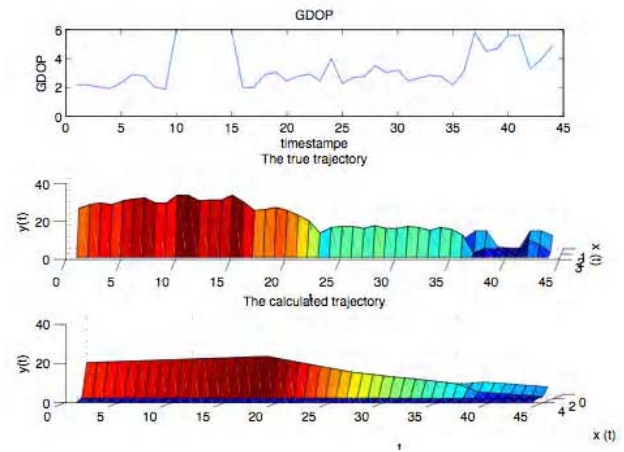

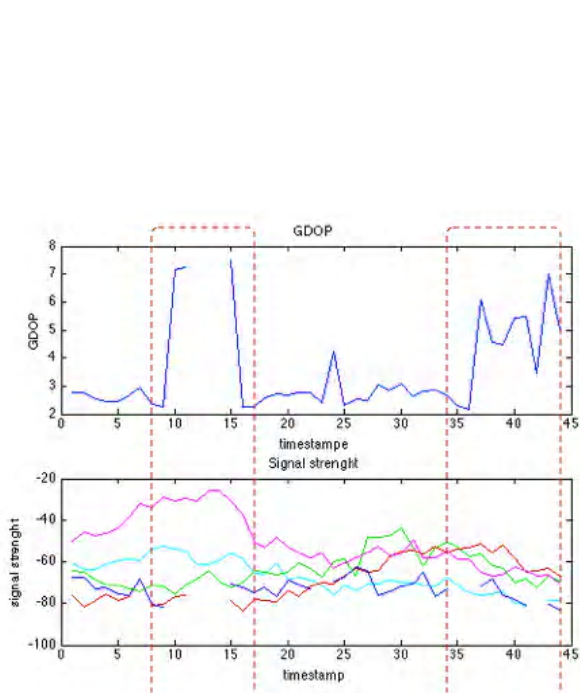

The third step of our experiments has been carried out to analyse the efficiency of our model by comparing the real trajectory and the estimated one with Wi-Fi DOP values (see Fig. 9). The analysis shows that the trajectories (the real one and the estimated one) are more or less similar except when the Wi-Fi DOP is up to eight.

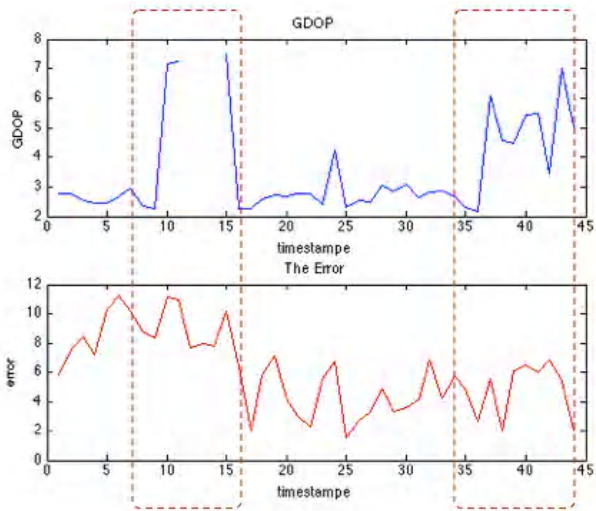

The fourth and last step is to verify whether the Wi-Fi DOP is a good indicator of the positioning accuracy. Fig.10 shows that when DOP [10,15], the error is up to eleven. When the Wi-Fi DOP value is equal to three (when the Wi-Fi DOP value is [1,5], we consider that the system has a god accuracy) the mean average error is equal to four.

## 6. Conclusions

Nowadays, the Wi-Fi positioning algorithms and systems are becoming a new mean of positioning mobile terminals within a heterogeneous environment.

The quality of service of such system may be improved in order to guaranty the integrity and the continuity of service.

This paper describes a model for dilution of precision and a mathematical description of the coefficient weakening of the accuracy, the Wi-Fi DOP.

The model presented in this paper may provide the guaranty we need. In fact, as shown in the results obtained in the previous section, our model illustrates the positioning system accuracy.

The idea consists in the observation of the results of the model and when the values of this one reach a given threshold, we inform the user that the position accuracy is not sufficient and then anticipate a solution to guaranty the quality and continuity of service.

## 7. Future Trends

Our model opens and leads to numerous extensions and perspectives.

The coefficient of dilution of precision or rather the Wi-Fi DOP is a good candidate to specify the most adequate access points distribution. It is possible to extend the Wi-Fi DOP to the system OWLPS.

It could provide a continuity of positioning, but also assistance to the optimal positioning of access points. The aim of this study is to offer to the user most of the time four access points with a DOP of the order of 2 in sight.

### Acknowledgments

We thank all the reviews for their detailed feedback and suggestions specially Matteo Cypriani.

**Soumaya Zirari** was born in 1981. She received her diploma in engineering in 2006. She is preparing her Ph.D Thesis at the Computer Science Laboratory at the University of Franche-ComtŽ in France, to be defended the 1st semester of 2010. She is focusing on hybrid location-based services and service continuity.

**Dr Philippe Canalda** got M.Sc. and Ph.D. Degrees in computer science from the University of OrlŽans (France) in 1991 and 1997, respectively. He worked at INRIA Rocquencourt from 1991 to 1996 on the automatic generation of optimizing and parallel n-to-n cross-compilers. From 1996 to 1998, he worked as Research Engineer in the Associated Compiler Expert start-up factory at Amsterdam, The Netherlands. Then he worked 2 years at LORIA on the synchronisation of cooperative process fragment, based on workflow model, and applied to ephemeral enterprise. Since 2001, he is an Associate Professor at the Computer Science Laboratory (LIFC, EA 4269) at the University of Franche-Comté in France. His research topics deal with, on the one hand mobility services and wireless positioning, and on the other hand on robust and flexible optimizing algorithms based on graph, automata and rewriting theories..

**Prof. François Spies** received his Ph.D. and the French "Accreditation to supervise research" Degrees in 1994 and 1999, respectively. He was an Associate Professor at the Computer Science Laboratory at the University of Franche-Comté in France from 1996-1999. Since 1999, he has held a Professor position at the University of Franche-Comté. Currently he is focusing on managing video streams on wireless and mobile architecture. Researches on, cooperative video cache strategies including mobility and video quality levels, transport, congestion control and quality of service of video streams are the main developed topics.

IJCSI